\documentclass[%
reprint,
superscriptaddress,
amsmath,amssymb,
aps,
pra,
]{revtex4-2}

\usepackage{graphicx}
\usepackage{dcolumn}
\usepackage{bm}
\usepackage{hyperref}
\usepackage{color}
\usepackage{amsfonts}
\usepackage{amsmath}
\usepackage{cancel}

\newcommand{\dd}{\mathrm{d}}
\newcommand{\Rabbitt}{RABITT}

\newcommand{\affcuni}{\affiliation{Institute of Theoretical Physics, Faculty of Mathematics and Physics, Charles University, V Holešovičkách 2, 180 00, Prague 8, Czech Republic}}

\newcommand{\affcnrs}{\affiliation{Universite Claude Bernard Lyon 1, CNRS, Institut Lumi\`ere Mati\`ere, UMR5306, F-69100, Villeurbanne, France}}

\begin{document}

\preprint{APS/123-QED}

\title{%
Angular momentum dependence in multiphoton ionization and attosecond time delays
}

\author{Jakub Benda}
\email{jakub.benda@matfyz.cuni.cz}
\affcuni

\author{Zdeněk Mašín}
\affcuni

\author{Sreelakshmi Palakkal}
\affcnrs

\author{Franck L\'epine}
\affcnrs

\author{Saikat Nandi}
\affcnrs

\author{Vincent Loriot}
\affcnrs

\date{\today}

\begin{abstract}
Multi-photon interference measurements in attosecond physics, e.g.\ the reconstruction of attosecond beating by interference of two-photon transitions (\Rabbitt{}), are the typical methods of choice to experimenally access the photoionization delay in atoms and molecules. The exact relation between the measurable multi-photon delays and the theoretical single-photon delays is typically modelled by correction terms, `continuum-continuum delays', obtained from a high-energy limit of the theory. However, these are unreliable at photoelectron kinetic energies smaller than about 10~eV, and do not have photoemission angular dependence. In this work we develop an accurate analytic alternative that gives accurate correction terms even at very low energies. Our method is computationally very straightforward, predicts correct multi-photon photoelectron angular distributions as well as the expected angular dependence of the continuum-continuum delay. We validate the approach theoretically against state-of-the-art \textit{ab initio} calculations as well as experimentally with a two-harmonic \Rabbitt{} setup, which properly separates different higher-order multi-photon pathways and offers a promising way of analysing congested molecular photoionization spectra.

\end{abstract}

\maketitle

\section{\label{sec:intro}Introduction}

An electron ionized by a single photon from an atom or a molecule scatters in the parent potential before reaching asymptotic distances. 
This affects the phase of the outgoing photoelectron, and can be semiclassically understood as a time delay \(\tau_W\) in the scattering process as initially proposed by Wigner \cite{WignerIonizTime, RMP2015}.
Ionization time delay thus probes the short-range potential landscape around the residual ion and as a new observable it has recently attracted a substantial amount of interests from the attosecond community and is now investigated in molecules and quantum systems of increasing sizes \cite{Boyer2023, Gong2022WaterCluster, LoriotNPhys24}.

The \Rabbitt{} protocol is one of the common methods to characterize an attosecond pulse train \cite{Paul1689} and to measure the ionization time delay with spectral resolution \cite{ScienceLHuillier2017}. In the latter case it produces an intuitive signal with a close correspondence to the Wigner ionization time delay. 
However, the measurement is based on interference of a pair of two-photon ionization pathways; that is, apart from the ionizing extreme ultra-violet (XUV) field, the system also absorbs or emits a quantum of another, infra-red (IR) dressing field. Consequently, the temporal \Rabbitt{} observable \(\tau_R\), ``\Rabbitt{} sideband delay'', is affected by the dressing field and cannot be directly linked to the intrinsic one-photon ionization delay.
Different strategies have emerged to isolate the Wigner time delay \(\tau_W\) from \Rabbitt{} measurements such as performing the full two-photon calculation~\cite{Mauritsson2005}. Up to date, though, the most widely used solution is based on a universal dressing field delay through the so-called `continuum-continuum' delay, $\tau_{cc}$, introduced for atoms by Dahlström \textit{et al.}~\cite{Dahlstrom2012}, so that \(\tau_R \approx \tau_W + \tau_{cc}\). 
The model has further been extended to molecules~\cite{WornerThITD, Heuser2016, Cirelli2018}.
In addition to $\tau_{cc}$, other correction terms have also to be considered if further physics needs to be accounted for,   such as the  `(ion) coupling delay' \(\tau_{\text{coupl}}\) in systems with large transition dipole moments or the `dipole-laser coupling delay' \(\tau_{dLC}\) in molecules with a sizeable permanent dipole \cite{Benda2022,Benda2024}. A common denominator of all these additional delays is that they are well defined only in the high kinetic energy limit where the photoelectron interacts with the residual ion only very weakly. At low energies the problem becomes more complex due to electron correlation, which effectively intertwines the effects responsible for all these asymptotic contributions.

In the present work we establish a computationally efficient and physically transparent approach to modelling and analysis of \Rabbitt{} delays for weakly correlated molecules. We generalize the idea of the universal continuum-continuum delay, and apply an IR correction \textit{separately} for each photoelectron partial wave, reflecting different influence of the IR field on individual angular momentum states. As a result, we are able to recover the correct low-energy behaviour of both the emission-integrated and angle-resolved \Rabbitt{} sideband delays, in excellent agreement with the full above-threshold perturbation theory of the second order. This makes it a most attractive tool for the attosecond community for routine interpretation of standard \Rabbitt{} experiments. Additionally, at a slightly reduced level of accuracy, the same correction factors can be used to calculate ionization amplitudes also for higher-order multi-photon ionization or for coupled models, paving a way for interpretation of more complex systems.

This paper presents complementary experimental and theoretical advances. On the experiment side, we employ the simplest \Rabbitt{} configuration conceivable, yet highly useful, that combines just one pair of adjacent odd harmonics of the fundamental field. The advantage of such arrangement is two-fold. 
First, by avoiding the complex cross-talk of different harmonics we make it possible to  measure the outcome of the higher-order processes that are otherwise responsible for hard-to-analyze contamination of the pristine two-photon \Rabbitt{} signal.
Second, it allows reading off of the oscillation from the interference bands forming elsewhere than just in between the two harmonics. This method is also advantageous at low intensity to reduce spectral congestion in molecules, where electronic states are closely spaced.

On the theory side, we model both the standard and the high-order \Rabbitt{} delays using the time-independent molecular above-threshold multiphoton R-matrix method~\cite{Benda2021} at the leading-order level of the perturbation theory, as well as using fully time-dependent simulations using `R-matrix with time-dependence' (RMT)~\cite{RMT}. This combination of state-of-the-art \textit{ab initio} methods allows us to validate and benchmark the newly proposed simple analytical solution for treating the dynamics of the photoionized system driven by the infrared (IR) field. 
Our method avoids the integration in the complex plane used e.g.\ in~\cite{Peschel2022} as well as the numerical quadrature over Green's kernel as in~\cite{Argenti_CK}, and provides a computationally cheap and accurate alternative to more sophisticated multiphoton \textit{ab initio}  calculations~\cite{Benda2022} and fully time-dependent simulations~\cite{Bertolino2021, Jiang2022, Bharti2023}. A key element of the present approach is inclusion of the angular momentum dependency of the ionization amplitudes affected by the dressing field. Such a property has already been experimentally observed~\cite{Fuchs2020,han24} and basic theoretical aspects of the angular momentum dependence have been discussed for hydrogen-like atoms~\cite{Ji_2024_cc},
within the WKB approximation in atomic systems~\cite{WKBcc} and
simulated in numerical calculations~\cite{ivanov2017, busto2019}. However, in the present work we generalize the theory to molecules, systems with transition dipoles, wide energy range, and to higher-order multi-photon processes.

The article is organized as follows.
Section~\ref{sect:derivation} presents the \Rabbitt{} interference pathways involving different orders of the perturbation theory. The Wigner time delay is derived and the asymptotic approximation of two-photon transitions is recalled. Our implementation of the partial-wave-resolved continuum-continuum contribution is introduced and generalized to arbitrary orders of interfering pathways for atoms and molecules.
Section~\ref{sect:results} gives numerical predictions  for angle-integrated and angle-resolved \Rabbitt{} delays. The results are illustrated with argon in detail and are compared with original measurements. We particularly investigate how the IR interaction modulates the angular distribution of \Rabbitt{} signal at different photon absorption orders and how the standard two-photon \Rabbitt{} sideband delay is reflected into higher-order interference bands.
The practical tutorial on how to introduce IR transitions to calculated partial wave-resolved XUV photoionization amplitudes is provided in Section~\ref{sect:practical}.

Additional applications of the new method are given in the Supplementary Material~\cite{supp}, such as the effective $\tau_{cc}$ extracted for the widely used noble gases, results for ionization from a molecular orbital of CO\textsubscript{2} that features a shape resonance, and the time delays in the polar molecule LiH.

\section{Method}
\label{sect:derivation}
\subsection{Interference pathways in \Rabbitt{}}
\label{sect:Multichannel}
\begin{figure}[!ht]
        \centering
        \includegraphics[width=8cm]{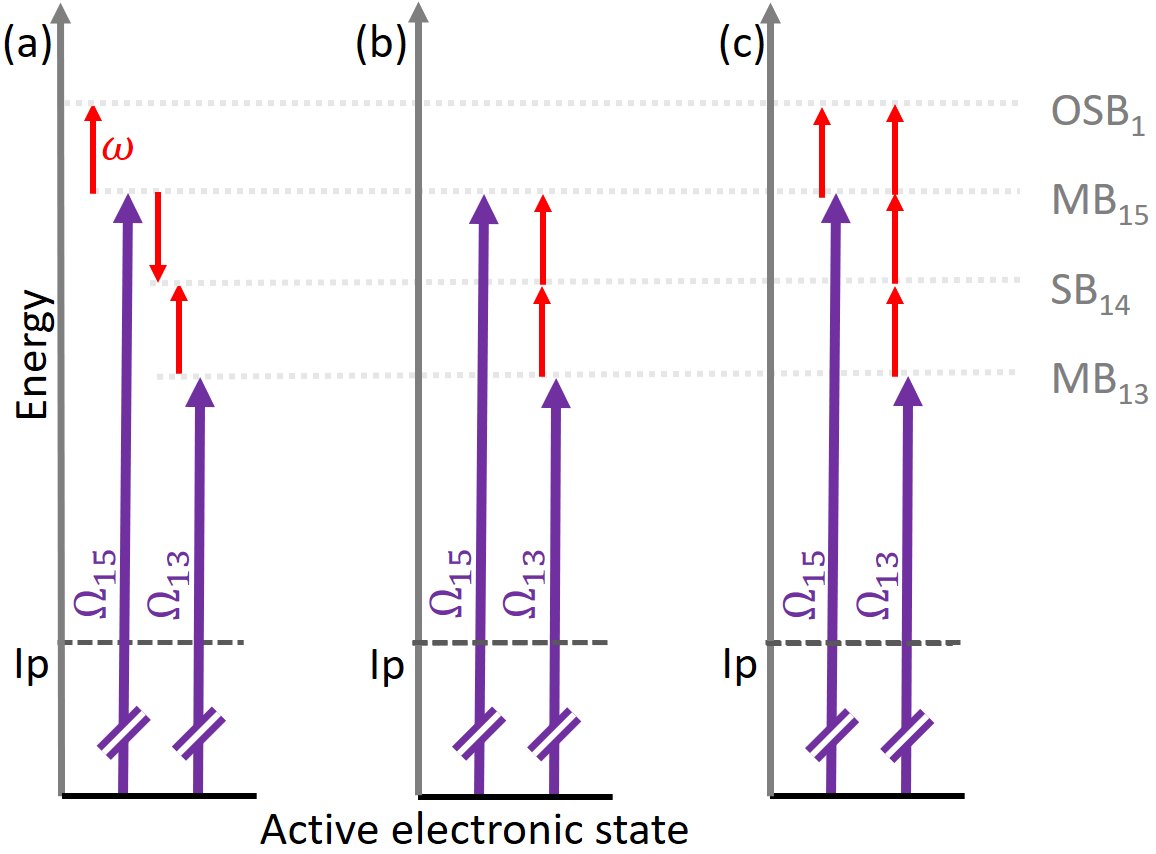}
        \caption{Pathways considering only two consecutive harmonics of the comb ($\Omega_{13}$ and $\Omega_{15}$) and a dressing field limited to (a) one, (b) two and (c) three photons absorbed or emitted reaching the sideband (SB), the `mainbands' (MB) or the upper outer sideband (OSB$_1$). Ip is the ionization potential.}
        \label{RabbittPrinciple}
    \end{figure}

In a \Rabbitt{} experiment, an attosecond pulse train is generated in a centrosymetric media that corresponds, in the spectral domain, to a comb of odd harmonics $\Omega$ reaching the XUV range, from a fundamental frequency $\omega$ (i.e. $\Omega=(2n+1)\omega$, for integer $n$). Each harmonic above the ionization potential (Ip) can ionise the target and a photoelectron can be emitted into the continuum. The photoelectron spectrum is hence composed of several well localised peaks (main bands (MB)) where the ionization phase information is experimentally inaccessible. 
To be sensitive to such phases, the \Rabbitt{} protocol proposes to dress the harmonic comb with a weak light pulse with a central frequency of $\omega$ interferometrically stabilised with the XUV pulse. The photoelectrons are hence redistributed into sidebands (SB) with an energy shift of $\pm\omega$. As a function of the dressing pulse intensity, several $\omega$-photons can be absorbed or emitted as shown in Fig.~\ref{RabbittPrinciple}. 
To unambiguously identify the interference pathways, a set of only two following harmonics is considered all along the manuscript ($\Omega_{13}$ and $\Omega_{15}$, i.e. separated by 2$\omega$) theoretically and experimentally. 
The addition of a single $\omega$-photon opens two quantum paths leading to the same final kinetic energy at SB$_{14}$ that interfere (see Fig.~\ref{RabbittPrinciple}(a)). 
By scanning the delay $\tau$ between the pulses, the SB intensity oscillates as
    \begin{eqnarray}
        I(\tau)&=& \mathcal{A}+\mathcal{B}\cos\left(2\omega\tau-\phi_{2\omega}\right)\nonumber \\
        &=& \mathcal{A}+\mathcal{B}\cos\left(2\omega(\tau-\tau_R)\right),        \label{RabbittFormula}
    \end{eqnarray}
with $\mathcal{A}$ being the baseline, $\mathcal{B}$ the  amplitude and $\phi_{2\omega}$ the phase of the oscillation at 2$\omega$, and $\tau_R=\phi_{2\omega}/(2\omega)$ the \Rabbitt{} delay. All of these quantities can be expressed using two-photon or higher-order dipole matrix elements between the initial neutral state and the final photoelectron state~\cite{ertel2024a}.
The information on the ionization time delay is stored within this \Rabbitt{} delay, which is also affected by the dressing field. 

\begin{figure}
    \centering
    \includegraphics[width=\columnwidth]{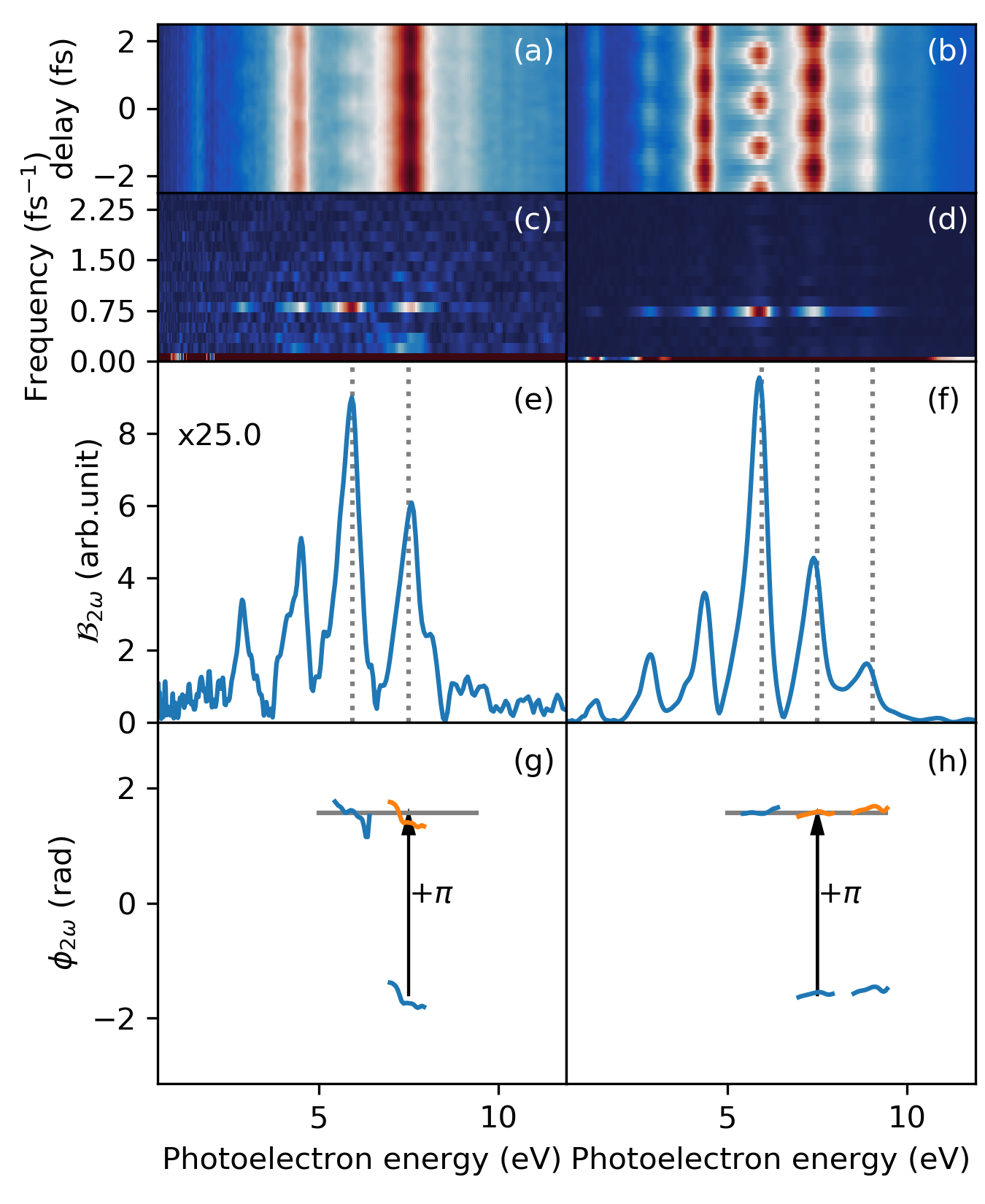}    
    \caption{(a, b) \Rabbitt{} spectrograms obtained in argon by using a set of two isolated harmonics ($\Omega_{13}$ and $\Omega_{15}$) for low (a) and high (b) dressing field intensity. (c,d) the corresponding Fourier transform showing a clear oscillation at 2$\omega$ (i.e. at 0.75~fs$^{-1}$). The 2$\omega$ oscillation amplitude (e,f) and phase (g,h) in blue. In (g,h), the phase of the MB and OSB are also presented in orange with a $\pi$-shift to be compared to the SB. Experimental details are in the supplementary material~\cite{supp}.}    \label{ExperimentRabbittAngularlyIntegrated}
\end{figure}

Since the dressing pulse is weak, it does not ionize the target but only redistributes the photoelectrons. The increase in the population of a SB implies the depopulation of its surrounding MBs. Since the amplitude of the SB depends on $\tau$, the MBs reflect the complementary behavior i.e. they oscillate following Eq.~\ref{RabbittFormula} but in phase opposition.
According to the perturbation theory, the MB oscillations are described by another interfering path.  
For instance, as shown in Fig.~\ref{RabbittPrinciple}(b) for MB$_{15}$ where the path ($\Omega_{15}$) interferes with $(\Omega_{13}+2\omega$). 
The SB and MBs oscillation amplitudes are both linear with the dressing field intensity despite they involve different orders of the perturbation theory. 
This is experimentally illustrated on the left panels of Fig.~\ref{ExperimentRabbittAngularlyIntegrated} performed at low dressing field intensity with two isolated harmonics (see Section II of the Supplementary Material~\cite{supp} for experimental details). MB$_{15}$ can be reached also by a beyond-leading-order transition $(\Omega_{15}+\omega-\omega$), which combines with the plain one-photon transition driven by $\Omega_{15}$. However, for weak dressing fields, such pathways are  suppressed with respect to the lower-order transition and can be neglected. Generalization to strong-field processes (\(\gtrsim 0.5\)~TW/cm\(^2\)) is possible within the time-dependent approach or the strong-field approximation~\cite{TD-RABITT}.

At low dressing field intensity, the first outer sideband (OSB1) appears through the pathway ($\Omega_{15}+\omega$) but it does not oscillate with $\tau$ because of the lack of an interfering pathway (see Fig.~\ref{RabbittPrinciple}(a) and Fig.~\ref{ExperimentRabbittAngularlyIntegrated}(left)). 
At higher dressing field intensity the $(\Omega_{13}+3\omega)$ transition becomes non-negligible and interferes with ($\Omega_{15}+\omega$) as shown in see Fig.~\ref{RabbittPrinciple}(c) and Fig.~\ref{ExperimentRabbittAngularlyIntegrated}(right). 
OSB1 also oscillates with $\tau$ following Eq.~\eqref{RabbittFormula} in phase to the MB and in phase opposition with the SB.  
The SB-MB-OSB coupled oscillations are demonstrated under the soft photon approximation in the Supplementary Material~\cite{supp} and using Floquet formalism in \cite{Lucchini2023Floquet}.
Let's notice that even at high dressing field intensity no 4$\omega$ oscillation (i.e. 1.5~fs$^{-1}$ for an 800~nm fundamental field) is observed since no harmonics are separated by 4$\omega$. 

All such inteference pathways depend on the XUV ionization phases that carry the ionization time delay. 
This introduces a redundancy in the information that can be exploited to extract the value with high accuracy. 
However, the dressing field affects  both the yield and angular distribution of the oscillation depending on the considered band. 
In the following, after expressing the Wigner time delay, a general expression is derived that accurately treats the influence of the dressing pulse on the SB, MB and OSB$n$.

\subsection{The Wigner time delay}
\label{sect:XUVOnly}
All along the analytical derivation, Hartree atomic units (\(\hbar = e = m_e = 4\pi \varepsilon_0 = 1\)) are used  unless another units are explicitly indicated. 
The formalism is based on the laser-assisted photoionization perturbation theory of Dahlström \textit{et al.}~\cite{Dahlstrom2012} with monochromatic fields. Here, only the necessary steps are recalled following the conventions defined in~\cite{Benda2024}. 

The one-photon ionization amplitude
\begin{align}
    T_\textrm{fi}^{(1)} = 2\pi i \langle \Psi_{\textrm{f}\bm{k}}^{(-)} | \hat{\bm{\varepsilon}}_\text{XUV} \cdot \bm{D} | \Psi_\textrm{i} \rangle
    = 2\pi i d_\textrm{fi}^{(1)}(\bm{k})
    \label{eq:T1}
\end{align}
from the initial bound state \(\Psi_\textrm{i}\) to a final continuum state \(\Psi_{\textrm{f}\bm{k}}^{(-)}\) for field with the polarization vector \(\hat{\bm{\varepsilon}}_\text{XUV}\) is expressed in terms of the one-photon ionization dipole matrix element
\begin{align}
    d_\textrm{fi}^{(1)}(\bm{k})
    = \sum_{lm} d_{\textrm{f}lm,\textrm{i}}^{(1)}(k) Y_l^m(\hat{\bm{k}}) \,,
    \label{eq:dfi}
\end{align}
where \(\bm{D}\) is the operator of the total electronic dipole moment, \(\bm{k}\) is the photoelectron momentum, and \(Y_l^m\) are the spherical harmonics.
After a single ionization by an XUV radiation, and at large distances of the ejected photoelectron from the molecule, the state of the system can be asymptotically written as the partial wave channel expansion~\cite{Benda2021}
\begin{equation}
    \Psi_{\textrm{i} + \Omega}^{(+)}(\bm{r}) \rightarrow \frac{1}{r} \sum_{\nu\lambda\mu} a_{\nu\lambda\mu,\text{i}}^{(1)} H_{\lambda}^+(-\tfrac{Z}{\kappa_\nu},\kappa_\nu r) Y_{\lambda}^{\mu}(\bm{r}) \Phi_{\nu} \,,
    \label{eq:intermasy}
\end{equation}
with \(H_l^\pm(\eta,\rho)\) a travelling Coulomb wave function, denoted in this paper also as \(H_l^\pm(r)\) for brevity. 
Asymptotically, its behavior is that of a Coulomb-corrected plane wave,
\begin{align}
    H_{\lambda}^+
    \rightarrow \exp \left[i\left(\kappa_\nu r + \frac{Z}{\kappa_\nu}\ln 2\kappa_\nu r - \frac{\pi\lambda}{2} + \sigma_{\lambda}(\kappa_\nu)\right)\right] \,.
    \label{eq:Hasy}
\end{align}
The symbol \(\Phi_{\nu}\) in Eq.~\eqref{eq:intermasy} stands for the electronic state of the residual ion coupled to the partial wave \(\lambda\mu\), and \(\sigma_\lambda(\kappa) = \arg \Gamma(\lambda + 1 - Zi/\kappa) \) is the Coulomb phase for a center with the residual charge \(Z\). In the equations  only the coordinates of the photoelectron are explicitly labeled. The expansion coefficient \(a_{\nu\lambda\mu,\text{i}}^{(1)}\) in Eq.~\eqref{eq:intermasy} is proportional to a specific partial wave component of the one-photon transition dipole~\cite{Benda2022},
\begin{align}
    a_{\nu\lambda\mu,\text{i}}^{(1)}
    &= -2\pi \int i^{\lambda} e^{-i\sigma_\lambda} Y_{\lambda}^{\mu*}(\hat{\bm{\kappa}}_\nu) \langle \Psi_{\nu\bm{\kappa}_\nu}^{(-)} | V_{\text{XUV}} | \Psi_\textrm{i} \rangle \dd^2\hat{\bm{\kappa}}_\nu \nonumber \\
    &= -\sqrt{\frac{2\pi}{\kappa_\nu}} i^{\lambda} e^{-i\sigma_\lambda} d_{\nu \lambda \mu, \textrm{i}}^{(1)}(\kappa_\nu) \,.
\end{align}
Here \(V_{\text{XUV}} = (\bm{D}_{\text{ion}} + \bm{r}) \cdot \hat{\bm{\varepsilon}}_{\text{XUV}}\) stands for the projection of the total electronic dipole operator along the polarization \(\hat{\bm{\varepsilon}}_{\text{XUV}}\) of the XUV field.

Following the Wigner-theory~\cite{WignerIonizTime}, the ionization time delay of a given partial wave $lm$ for an electron energy \(E_k = k^2/2=\Omega-\text{Ip}\) corresponds to the derivative of the phase of an XUV transition dipole with respect to energy,
\begin{equation}
    \tau_{W,lm}(E_k)=\frac{\partial}{\partial \Omega} \arg \left (d^{(1)}_{\textrm{f}lm,\textrm{i}}(\Omega)\right).
\end{equation}
In most cases, the final state corresponds to a partial-wave mixture and the ionization time delay depends on the emission angle. Within the \Rabbitt{} framework the differentiation step becomes the energy gap between two adjacent odd harmonics ($\partial\Omega\rightarrow 2\omega$). For oriented emission of a photoelectron at energy \(E_k = k^2/2\), one then recovers the `atomic' one-photon delay
\begin{equation}
    \tau_1(\bm{k}) = \frac{1}{2\omega} \arg 
    \left( d_{\textrm{fi}}^{(1)*}(\kappa_+\hat{\bm{k}})
    d_{\textrm{fi}}^{(1)}(\kappa_-\hat{\bm{k}})\right) \,,
\end{equation}
where \(\hat{\bm{k}}\) is a unit vector, and \( \kappa_\pm = \sqrt{2(E_k \mp \omega)}\) is the intermediate momentum in the \((\Omega_<+\omega)\) and \((\Omega_>-\omega)\) pathways, respectively. This quantity represents a finite difference approximation to an effective Wigner delay and  corresponds to the expected value to extract from the \Rabbitt{} protocol. All along the manuscript $\tau_1$ is taken as the reference for oriented emission. For emission-integrated and orientation-averaged \Rabbitt{} signal we use instead the reference formula
\begin{align}
    \tau_1'(k) = \frac{1}{2\omega} \arg \sum_{l'm'lm ab}
    \langle l'm' | \hat{n}_a \hat{n}_b | lm \rangle \nonumber \\
    \times \int \hat{\varepsilon}_a \hat{\varepsilon}_b 
    d_{\textrm{f}l'm',\textrm{i}}^{(1)*}(\kappa_+)
    d_{\textrm{f}lm,\textrm{i}}^{(1)}(\kappa_-) \dd^2 \hat{\bm{\varepsilon}} \,,
\end{align}
which corresponds to the `molecular' delay~\cite{WornerThITD}; \(\hat{\bm{\varepsilon}}\) denotes the common linear polarization vector of XUV and IR.

\subsection{The asymptotic theory}
\label{sect:asystd}
In this section, we review the widely used asymptotic theory  that separates the ionization time delay from the measurement-induced time shift due to the dressing field.
The original formulation has been developed for atoms by Dahlström \textit{et al.}~\cite{Dahlstrom2012}, extended to molecules by Baykusheva and Wörner~\cite{WornerThITD}, and its application to  systems with coupled channels was  further discussed \cite{Kamalov2020,Benda2022}.

The two-photon ionization amplitude is calculated as the dipole transition between the intermediate state and a proper final stationary photoionization state \(\Psi_{\textrm{f}\bm{k}}^{(-)}\) given by the boundary condition~\cite{BurkeRM}
\begin{gather}
    \Psi_{\textrm{f}\bm{k}}^{(-)}(\bm{r}) \rightarrow \sum_{lm} i^l e^{-i\sigma_l(k)} Y_{l}^{m*}(\hat{\bm{k}})
    \sum_{g\lambda\mu} F_{\textrm{f}lm,g\lambda\mu}^{(-)}(r) Y_{\lambda}^{\mu}(\hat{\bm{r}}) \Phi_{g} \,, \nonumber \\
    F_{\textrm{f}lm,g\lambda\mu}^{(-)}(r) = \frac{-i}{r\sqrt{2\pi k}}(H_l^+ \delta^{\textrm{f}lm}_{g\lambda\mu} - H_l^- S_{g\lambda\mu}^{\textrm{f}lm*}) \,.
    \label{eq:finalasy}
\end{gather}
For pure one-electron Coulomb problem the \(S\)-matrix is trivial, the formula applies to all distances, and it simplifies to the regular Coulomb wave \(F_l(\eta, \rho)\) in each partial wave channel:
\begin{gather}
    \Psi_{\textrm{f}\bm{k}}^{(-)}(\bm{r}) = \sum_{lm} i^l e^{-i\sigma_l(k)} Y_{l}^{m*}(\hat{\bm{k}})
    Y_{l}^{m}(\hat{\bm{r}}) F_{\textrm{f}lm}^{(-)}(r) \Phi_\textrm{f} \,, \nonumber \\
    F_{\textrm{f}lm}^{(-)}(r) = \frac{1}{r} \sqrt{\frac{2}{\pi k}} F_l(-\tfrac{Z}{k}, kr) \,.
    \label{eq:finalhydro}
\end{gather}
The asymptotic approximation assumes validity of the asymptotic forms of the wavefunctions  Eq.~\eqref{eq:intermasy}  and~\eqref{eq:finalasy} throughout the whole radial range including the origin and neglects the non-diagonal, \(S\)-matrix-dependent second term in Eq.~\eqref{eq:finalasy}. 

Another possibility, employed in this work, is to replace the asymptotic form of the final-state wavefunction Eq.~\eqref{eq:finalasy} with an exact hydrogenic solution~Eq.~\eqref{eq:finalhydro}.
This will prove advantageous later, because the regular Coulomb function from Eq.~\eqref{eq:finalhydro} has the correct angular-momentum-dependent asymptotics \(r^{l + 1}\) at \(r \rightarrow 0\) and does not diverge as opposed to the usual choice Eq.~\eqref{eq:finalasy}.

Ultimately, the resulting two-photon amplitude is written as~\cite{Faisal}
\begin{align}
    T_\textrm{fi}^{(2)}(\bm{k})
    &= -2\pi i \langle \Psi_{\textrm{f}\bm{k}}^{(-)} | V_{\text{IR}} | \Psi_{\textrm{i}+\Omega}^{(+)} \rangle \nonumber \\
    &= \sum_{l m} T_{\textrm{fi},lm}^{(2)}(k) Y_l^m(\hat{\bm{k}}) \,,
    \label{eq:T2}
\end{align}
where each partial wave channel contribution
\begin{equation}
    T_{\textrm{fi},lm}^{(2)}(k) \approx \sum_p [T_{\textrm{fi},lmp,\text{pws}}^{(2)}(k) + T_{\textrm{fi},lmp,\text{ion}}^{(2)}(k)],
    \label{eq:T2pw}
\end{equation}
arises in a free-free or ion-ion transition:
\begin{align}
    T_{\textrm{fi},lmp,\text{pws}}^{(2)}(k) &= - 2\pi i\langle l m | \hat{\bm{\varepsilon}}_{\text{IR}} \cdot \hat{\bm{r}} | \lambda_p \mu_p \rangle \delta_\textrm{f}^{\nu_p} \nonumber \\
    &\qquad \times A_{\kappa_p \lambda_p k l}^{(1)} d_{\nu_p \lambda_p \mu_p,\textrm{i}}^{(1)}(\kappa_p) \,, \label{eq:T2pws}\\
    T_{\textrm{fi},lmp,\text{ion}}^{(2)}(k) &= - 2\pi i\langle \Phi_\textrm{f} | \hat{\bm{\varepsilon}}_{\text{IR}} \cdot \bm{D}_{\text{ion}} | \Phi_{\nu_p} \rangle \delta_{l}^{\lambda_p} \delta_{m}^{\mu_p} \nonumber \\
    &\qquad \times A_{\kappa_p \lambda_p k l}^{(0)} d_{\nu_p \lambda_p \mu_p,\textrm{i}}^{(1)}(\kappa_p) \,. \label{eq:T2ion}
\end{align}
The continuum-continuum integral \(A_{\kappa \lambda k l}^{(s)}\) is
\begin{align}
    A_{\kappa \lambda k l}^{(s)} &= -\frac{2}{\sqrt{\kappa k}} i^{\lambda - l} e^{i\sigma_l - i\sigma_{\lambda}} \nonumber \\
    &\quad \times \int_0^\infty F_l(-\tfrac{Z}{k}, k r) r^s H_{\lambda}^+(-\tfrac{Z}{\kappa}, \kappa r) \dd r \,.
    \label{eq:Tlli}
\end{align}
While the partial-wave mixing angular integrals in Eqs.~\eqref{eq:T2pws} and \eqref{eq:T2ion} have always been an integral ingredient of the asymptotic approximation and were discussed at length by Baykusheva and Wörner~\cite{WornerThITD}, the radial integration in Eq.~\eqref{eq:Tlli} has always been strongly approximated.
The integrand is usually simplified by neglecting the first term of \(F_l = (H_l^+ - H_l^-)/2i\) and replacing both remaining Coulomb-Hankel functions with their long-range asymptotics~\eqref{eq:Hasy}. 
In other words, it corresponds to an approximation \(A_{\kappa \lambda k l}^{(s)} \approx A_{\kappa k}^{(s)}\) that cancels the sensitivity to the intermediate and final partial waves, and leads to the expression
\begin{gather}
    A_{\kappa \lambda k l}^{(s)} \rightarrow A_{\kappa k}^{(s)} =
    i^{s} \frac{e^{-Z\pi/2\kappa + Z\pi/2k}}{\sqrt{\kappa k} (\kappa - k)^{s + 1}}
    \frac{\Gamma(s + 1 + \tfrac{Zi}{\kappa} - \tfrac{Zi}{k})}{(\kappa - k)^{Zi/\kappa - Zi/k}} \nonumber \\
    \times \frac{(2\kappa)^{Zi/\kappa}}{(2k)^{Zi/k}}
    \left[
    1 + \delta_{s1} \frac{Zi}{2}
    \frac{(\kappa^{-2} + k^{-2})(\kappa - k)}{1 + Zi/\kappa - Zi/k}
    \right]
    \,,
    \label{eq:Akk}
\end{gather}
that can be possibly further refined by
making other \textit{ad hoc} corrections (`regularized' variant).

\begin{figure}
    \centering
    \includegraphics[width=\columnwidth]{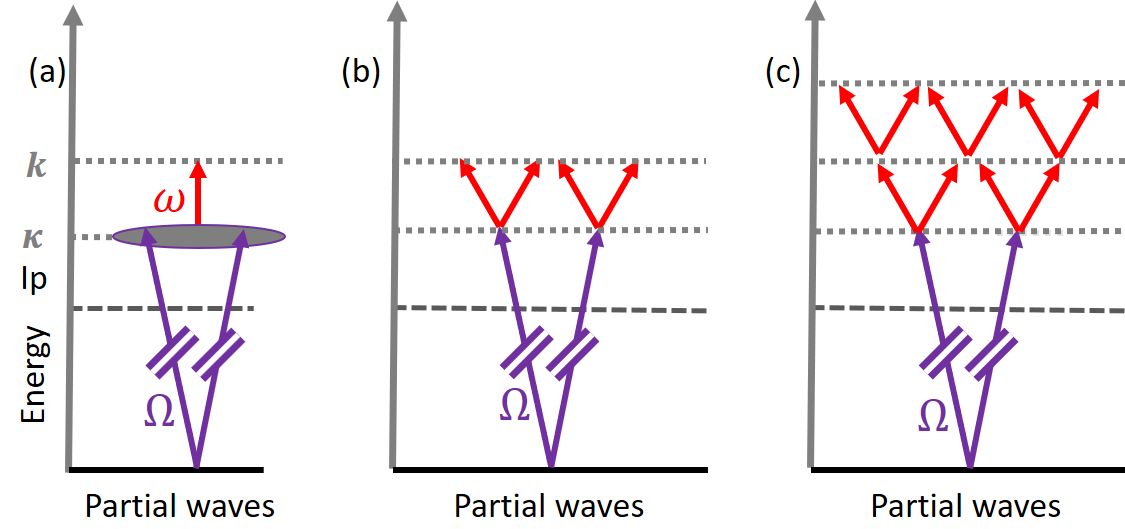}
    \caption{(a) 'Universal' dressing approach where the XUV and dressing photons behave independently.  Partial wave resolved approach (this work) where (b) one and (c) two $\omega$-photons transitions are considered.}
    \label{IonizationTimeDelayPrinciple}
\end{figure}

In absence of ion-ion transitions the replacement of the Coulomb-Hankel functions \(H^\pm\) by exponentials removes all partial-wave dependence from the infinite integral in Eq.~\eqref{eq:Tlli}. This allows to treat the factor as a `universal' correction independent of the specific system and partial wave,
\begin{equation}
    T_\textrm{fi}^{(2)}(\bm{k}) \approx -2\pi iA_{\kappa k}^{(1)} d_\textrm{fi}^{(1)}(\kappa \hat{\bm{k}}) \,,
    \label{eq:factT2}
\end{equation}
giving rise, ultimately, to the separability of the \Rabbitt{} delay into the universal direction-independent continuum-continuum delay \(\tau_{cc}\) and the one-photon (Wigner-like) delay \(\tau_1\),
\begin{align}
    \tau_R(\bm{k}) &= \frac{1}{2\omega} \arg (T_{\textrm{fi},+}^{(2)*} T_{\textrm{fi},-}^{(2)}) \nonumber \\
    &\approx \frac{1}{2\omega} \left[ \arg (A_{\kappa_+ k}^{(1)*} A_{\kappa_- k}^{(1)})
    + \arg (d_\textrm{fi}^{(1)*}(\hat{\bm{k}}\kappa_+) d_{\textrm{fi}}^{(1)}(\hat{\bm{k}}\kappa_-))\right] \nonumber \\
    \tau_R(\bm{k})&\approx \tau_{cc}(k) + \tau_1(\bm{k}) \,.
    \label{eq:tauRsep}
\end{align}
This decomposition in two independent steps is illustrated in Fig.~\ref{IonizationTimeDelayPrinciple}(a) where the dressing pulse affects all photoelectron partial waves equally.

\subsection{Partial wave dependent IR contribution}
\label{sect:pwasy}

The crucial point of our work is to include the coupling of the photoelectron angular momentum to the dressing field. We do this by evaluating the continuum-continuum transition amplitude Eq.~\eqref{eq:Tlli} \textit{exactly}. Use of the exact partial-wave specific correction factors \(A_{\kappa\lambda kl}\) introduces a qualitative change: it generally prevents the factorization of the amplitudes given by Eq.~\eqref{eq:factT2} and the separability of the delays given by Eq.~\eqref{eq:tauRsep}. Only in simple systems, where there is only one relevant residual ion state (f$_\circ$) and one intermediate partial wave \((\lambda_\circ,\mu_\circ)\), one can still achieve some factorization, as then the summation over the contributions of the intermediate partial waves in Eq.~\eqref{eq:T2pw} reduces to a single term and the ion-ion transition does not contribute,
\begin{align}
    T_\textrm{f$_\circ$i}^{\circ{}(2)}(\bm{k}) &= -2\pi i\left[ \sum_{lm} A_{\kappa\lambda_\circ kl}^{(1)}
    \langle lm|\hat{\bm{\varepsilon}}\cdot\hat{\bm{r}}|\lambda_\circ \mu_\circ\rangle Y_l^m(\hat{\bm{k}}) \right] \nonumber \\
    &\qquad \times d_{\textrm{f}_\circ,\lambda_\circ\mu_\circ,\textrm{i}}^{(1)}(\kappa) \,.
\end{align}
This factorization leads to separation of the \Rabbitt{} delay to
\begin{equation}
    \tau_R^\circ(\bm{k}) \approx \tau_{cc}(\bm{k}) + \tau_1^\circ(k) \,.
\end{equation}
In contrast to the conventional situation in Eq.~\eqref{eq:tauRsep}, here the one-photon delay \(\tau_1^\circ\) is angularly-independent because the intermediate wavefunction consists of a single partial wave only, but it becomes angularly-dependent  through the continuum-continuum delay \(\tau_{cc}\). In less trivial systems the combination is more complex and becomes difficult to interpret as a whole. Instead, one has to work with individual partial waves composing the total ionization signal as given by Eqs.~\eqref{eq:T2pw}--\eqref{eq:T2ion}.

To sum up, the asymptotic approximation used to analyze the \Rabbitt{} experiments can be qualitatively significantly improved  by calculating the \(A_{\kappa\lambda kl}\) coefficients more accurately. This naturally leads to distinct photoionization delays accumulated by individual partial waves in the free-free transition. Consequently, such approach allows improved interpretation of angular dependence and angular averaging of \Rabbitt{} oscillation, where multiple non-equivalent partial waves mix and interact.
Also, neither use of \textit{ab initio} time-dependent method, nor of the second order perturbation theory are needed to obtain the amplitudes: the only input data are the one-photon ionization amplitudes.

While there is no closed-form expression for the value of the integral in Eq.~\eqref{eq:Tlli}, its numerical evaluation can be done at rather low computational cost. Additionally, it is a \textit{universal} coefficient that can be tabulated once and for all and used for any target. In our implementation the integration range is divided into two regions as follows:
\begin{itemize}
\item The region \(r < r_1\) before some asymptotic distance \(r_1\) is integrated using Levin quadrature~\cite{Levin}, which is well-suited for highly oscillatory functions and has proved invaluable for the R-matrix multi-photon method~\cite{Benda2022}.
\item The asymptotic region \(r_1 < r < +\infty\) is treated by regularization using a decreasing exponential, expansion of both Coulomb functions in asymptotic series and by integrating their product term by term using asymptotic integration by parts~\cite{AymarCrance,Benda2021}. This is similar to the regularization technique in the recent article~\cite{Ji_2024_cc}, though in the present method the damping coefficient can be always set to zero before evaluating the asymptotic formulas numerically.
\end{itemize}
The parameter \(r_1\) is subject to convergence checking, but we have found the value of \(r_1 = 100\)~a.u.\ to give good agreement with direct second-order perturbation theory calculations demonstrated in Section~\ref{sect:results}. 
Additionally, the special case of the radial integral in Eq.~\eqref{eq:Tlli} for the ion-ion transitions, with \(s = 0\) and \(l = \lambda\), can be also calculated analytically and expressed in a closed form~\cite{CoulInteg}:
\begin{gather}
    \lim_{c \rightarrow 0^+} \int_0^\infty F_l(-\tfrac{Z}{k},kr) H_l^+(-\tfrac{Z}{\kappa},\kappa r) \mathrm{e}^{-cr} \dd r = \nonumber\\
    \frac{k}{k^2 - \kappa^2} \left(\frac{k}{\kappa}\right)^l
    e^{-Z\pi/2\kappa + Z\pi/2k} \left|\frac{\Gamma(l + 1 - Zi/k)}{\Gamma(l + 1 - Zi/\kappa)}\right| .
\end{gather}
Here the imaginary component of \(H_l^+\) does not contribute due to orthogonality of the regular Coulomb waves and the only contribution comes from the real part of \(H_l^+\). The formula has been obtained from the general one~\cite{CoulInteg} using the known limiting behaviour of Coulomb functions at \(r \rightarrow 0\). Substituting this identity into Eq.~\eqref{eq:Tlli} leads to
\begin{equation}
    A_{\kappa l kl}^{(0)} = \frac{2}{k^2 - \kappa^2}
    \frac{k^{l+1/2} e^{Z\pi/2k} \Gamma(l + 1 - Zi/k)}{\kappa^{l+1/2} e^{Z\pi/2\kappa} \Gamma(l + 1 - Zi/\kappa)} \,.
    \label{eq:Akk0}
\end{equation}

\begin{figure}[htbp]
    \centering
    \includegraphics[width=\columnwidth]{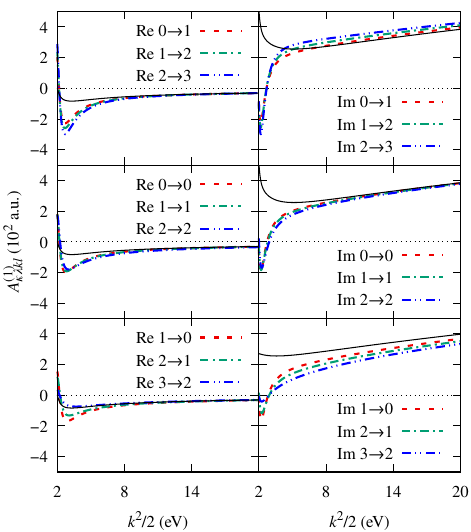}
    \caption{Values of \(A_{\kappa\lambda kl}^{(1)}\) for absorption of an 800~nm quantum, compared to the `long-range' variant of the traditional asymptotic approximation (\(A_{\kappa k}^{(1)}\), solid black). Left and right panels show the real and imaginary part, respectively, for selected \(\lambda \rightarrow l\) angular momentum transitions as labelled in the key of the individual panels.}
    \label{fig:akkl}
\end{figure}

Fig.~\ref{fig:akkl} compares the calculated values of the coefficients \(A_{\kappa \lambda k l}^{(1)}\) to the traditional `long-range' asymptotic approximation \(A_{\kappa k}^{(1)}\) for some angular momentum transitions. 
At high energies the two approximations converge, but they differ significantly at low photoelectron energies. The evaluated Coulomb integrals for the plotted and many other transitions are included in the public repository~\cite{zenodo}.

\subsection{Higher-order interference pathways}
\label{sect:higher-orders}

The intermediate state \(\Psi_{\textrm{i}+\Omega}^{(+)}\), that represents the state of the system after absorption of an $\Omega$-photon, can be used to obtain further intermediate states corresponding to additional $\omega$-photon absorptions or emissions. For the case with no ion-ion transitions, this has been proposed by Bharti \textit{et al.}~\cite{Bharti2021}. Additional $\omega$ absorption can be included in the time-independent description by means of solution of the time-independent Schrödinger equation. The first iteration, leading to intermediate state that interacted once with $\Omega$ and once with $\omega$ is
\begin{equation}
    (E_\textrm{i} + \Omega + \omega - H)\Psi_{\textrm{i} + \Omega + \omega}^{(+)} = V_{\text{IR}} \Psi_{\textrm{i} + \omega}^{(+)} \,.
\end{equation}
Within an approximation of uncoupled ionization channels, this equation can be solved by application of the hydrogenic Coulomb-Green's function for the photoelectron,
\begin{align*}
    G^{(+)}(\bm{r}, \bm{r'}) &= \langle \bm{r} | \frac{1}{E_k - H + i0} | \bm{r}' \rangle \nonumber \\
    &= \frac{1}{rr'} \sum_{lm} g_l^{(+)}(r, r') Y_{lm}(\hat{\bm{r}}) Y_{lm}(\hat{\bm{r}}')^* \,,
\end{align*}
to all channels. The radial part of \(G\) is
\begin{equation}
    g_l^{(+)}(r, r') = -\frac{2}{k}F_l(r_<) H_l^+(r_>) \,.
    \label{eq:gl}
\end{equation}
The resulting second intermediate state for total energy \(E_{\text{tot}} = E_\textrm{i} + \Omega + \omega = e_\textrm{f} + k^2/2\) is
\begin{align}
    \Psi_{\textrm{i} + \Omega + \omega}^{(+)}(\bm{r})
    &= \int G^{(+)}(\bm{r}, \bm{r}') V_\text{IR}(\bm{r}') \Psi_{\textrm{i} + \Omega}^{(+)}(\bm{r}') \dd^3\bm{r}', \nonumber \\
    &= -\frac{2}{k} \sum_{lm} Y_{lm}(\hat{\bm{r}}) \Phi_\textrm{f} \langle lm | \hat{\bm{\varepsilon}}_{\text{IR}} \cdot \hat{\bm{r}} | \lambda\mu \rangle a_{\textrm{f}\lambda\mu,\text{i}}^{(1)} \nonumber \\
    &\quad \times \int_0^\infty F_l(r_<) H_l^+(r_>) r' H_\lambda^+(r') \dd r' \,.
    \label{eq:Psi3}
\end{align}
This can be inserted into the formula for the leading-order perturbation theory transition amplitude,
\begin{align}
    T_\textrm{fi}^{(3)}(\bm{k}) = 2\pi i \langle \Psi_{\textrm{f}\bm{k}}^{(-)} | V_{\text{IR}} | \Psi_{\textrm{i} + \Omega + \omega}^{(+)} \rangle \,,
    \label{eq:T3}
\end{align}
and simplified by keeping only such radial integrals that contain Coulomb-Hankel functions of complementary signs. The resulting expression for the third-order amplitude then features triangular radial integrals of the kind
\begin{align}
    \int_0^\infty F_L(r) r H_l^+(r) \int_0^r F_l(r') r' H_\lambda^+(r') \dd r' \dd r \,.
    \label{eq:int2cont}
\end{align}
This method can be further generalized to even higher orders shown in Fig.~\ref{RabbittPrinciple}, and to ion-ion transitions by accounting also for the ion-ion transition term \(\bm{D}_\text{ion}\cdot\hat{\bm{\varepsilon}}\) in \(V_\text{IR}\) that has been disregarded in Eq.~\eqref{eq:Psi3} for simplicity. Each additional absorption or emission of an $\omega$-photon thus adds another layer of nested integration. In~\cite{Bharti2021} these radial integrals were separated into individual factors, essentially by fixed identification of \(r\) and \(r'\) with \(r_<\) and \(r_>\), respectively, in Eq.~\eqref{eq:gl}. This is motivated by a similar factorization implicitly taking place in the standard \Rabbitt{},
\begin{align}
    \int_0^\infty F_L(r) r H_l^+(r) \int_0^r F_l(r') r' P_{\text{i}}(r') \dd r' \dd r
    \nonumber \\
    \approx
    \underbrace{\int_0^\infty F_L(r) r H_l^+(r) \dd r'}_{\rightarrow A_{\kappa l k L}^{(1)}}
    \underbrace{\int_0^\infty F_l(r') r' P_{\text{i}}(r') \dd r'}_{\rightarrow d_{l,\text{i}}^{(1)}}
    \,,
    \label{eq:int2bound}
\end{align}
where \(P_\text{i}(r)\) represents radial part of the initial (Dyson) orbital. Due to the short range of \(P_\text{i}(r)\), the factorization \eqref{eq:int2bound} is meaningful, because the extension of the integration domain contains effectively zeros. However, when Eq.~\eqref{eq:int2cont} is factorized in the same way, the integration domain is doubled even though the integrand is non-zero in the newly added integration space. This has to be compensated by an appropriate combinatoric factor. For an \(n\)-dimensional integral this factor is \((n!)^{-1}\), giving
\begin{align}
    \int\limits_0^\infty f_1 \int\limits_0^{r_1} f_2 \dots \int\limits_0^{r_{n-1}} f_n \dd r_1 \dots \dd r_n 
    \approx \frac{1}{n!} \prod_{j = 1}^n \int\limits_0^\infty f_j \dd r_j \,.
    \label{eq:intfac}
\end{align}
This factorial factor is missing in the derivation~\cite{Bharti2021}.

Consequently, even amplitudes for higher-order transitions can be approximated in a factorized way. However, every possible photoelectron angular momentum pathway (e.g., \(\lambda_p \rightarrow \lambda_q' \rightarrow l\) in case of ($\Omega+2\omega$) interactions) now needs a different set of factors representing the individual $\omega$-photon absorptions. For pure free-free transitions (in absence of ion-ion transitions) in ($\Omega+2\omega$) process this gives the second intermediate state
\begin{gather}
    \Psi_{\textrm{i}+\Omega+\omega}^{(+)}(\bm{r}) \approx
    \frac{1}{r} \sum_{pq} a_{\nu_p \lambda_p \mu_p,\text{i}}^{(1)}
    \langle \lambda_q' \mu_q' | \hat{\bm{\varepsilon}}_{\text{IR}} \cdot \hat{\bm{r}} | \lambda_p \mu_p \rangle
    Y_{\lambda_q'}^{\mu_q'}(\hat{\bm{r}}) \Phi_{\nu_p} \nonumber \\
    \times \left(-\frac{2}{\kappa_q'}\right)  H_{\lambda_q'}^+(r) \int_0^{r}
    F_{\lambda_q'}(r') r' H_{\lambda_p}^+(r') \dd r' ,
\end{gather}
and the leading-order perturbation amplitude
\begin{align}
    T_\textrm{fi}^{(3)}(\bm{k})
    &\approx \frac{1}{2!} 2\pi i\sum_{lmpq} 
    A_{\kappa_p \lambda_p \kappa_q' \lambda_q'}^{(1)} A_{\kappa_q' \lambda_q' k l}^{(1)} \nonumber \\
    &\quad \times \langle l m | \hat{\bm{\varepsilon}}_{\text{IR}} \cdot \hat{\bm{r}} | \lambda_q' \mu_q' \rangle
    \langle \lambda_q' \mu_q' | \hat{\bm{\varepsilon}}_{\text{IR}} \cdot \hat{\bm{r}} | \lambda_p \mu_p \rangle \nonumber \\
    &\quad \times d_{\textrm{f}\lambda_p\mu_p,\textrm{i}}^{(1)}(\kappa_p) Y_{l}^{m}(\hat{\bm{k}}) \,.
\end{align}

\begin{figure}[!htbp]
    \centering
    \includegraphics[width=\columnwidth]{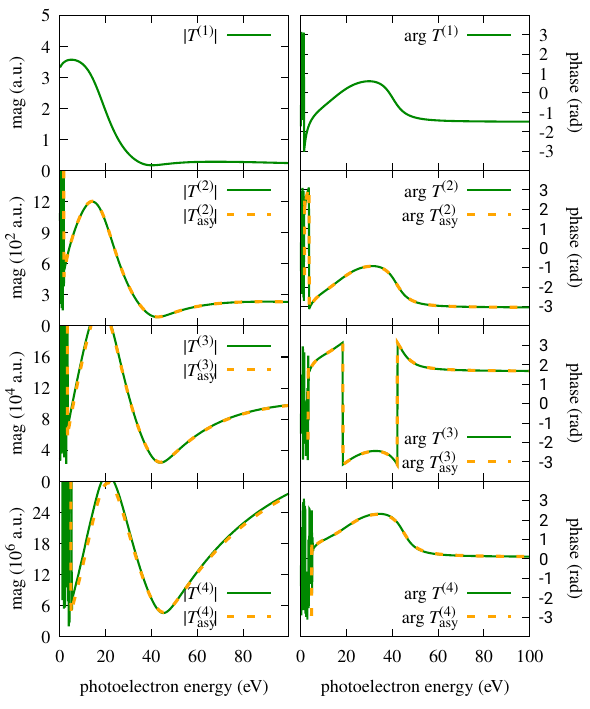}
    \caption{Multi-photon ($\Omega+n\omega$) ionization amplitudes of argon in static exchange model for  emission along the polarization axis of the 800~nm dressing field. Magnitudes (left panels) and phases (right panels) are plotted for (from top to bottom) \(n = 0, 1, 2, 3\). Solid lines: Amplitudes from the leading-order perturbation theory. Dashed lines: Partial-wave asymptotic approximation using first-order amplitudes.}
    \label{fig:ar-mag-arg}
\end{figure}

The agreement between the asymptotic approximation and a full multi-photon calculation for higher orders is compared in Fig.~\ref{fig:ar-mag-arg}. The calculations use a non-relativistic static exchange model of argon, where only the magnetic sublevel \(m = 0\)  of the \(3p\) shell is made available for ionization. The polarization and emission directions point along the \(z\) axis in this calculation. The \textit{ab initio} multi-photon ionization matrix elements used were obtained from UKRmol+~\cite{UKRmolp}. The approximate amplitudes agree very well with the multi-photon ones, including the magnitude, pointing to the importance of the permutation factor discussed above.

\section{Results}
\label{sect:results}

To illustrate the performance of the proposed method, we present results using the two harmonics \Rabbitt{} protocol that isolates the interferences and high orders are observed up to 3$\omega$ dressing photons. The theoretical results are supported by measurements in argon with angular resolution. 
Ionization time delays along the polarization axis and angularly integrated \Rabbitt{} delays are shown in at low photoelectron energy where the asymptotic approximations is becoming crucial and also calculated at larger photon energy. 
Additional results, including angular distribution
in argon, as well as delays for other atomic gases and in molecules demonstrating utility of the partial wave asymptotic theory all the way down to the boundary of the under-threshold \Rabbitt{} area \cite{Kheifets_2023} are provided in the Supplementary Material~\cite{supp}.

The calculation, performed with UKRmol+~\cite{UKRmolp} in \(D_{2h}\) point group as the largest available group, is based on a static-exchange model with Hartree-Fock (HF) orbitals of neutral argon obtained in Psi4~\cite{PSI4} from cc-pVDZ basis. It involves three symmetry components \(B_{3u}\), \(B_{2u}\) and \(B_{1u}\) of the highly symmetric ground state of Ar\(^+(3p^{-1})\) and radial continuum basis set consisting of B-splines spanning the radial range from the origin to the R-matrix radius \(r_a = 150\,a_0\) (Bohr radii). The wavelength of the dressing field is assumed \(\lambda = 800\)~nm.
The two-photon \Rabbitt{} delay \(\tau_R\) is calculated by the molecular multi-photon above-threshold ionization method~\cite{Benda2022}. The asymptotic approximation to the latter is obtained from Eqs.~\eqref{eq:T2pw} and~\eqref{eq:T2pws}, reusing the one-photon XUV-only dipoles.  The continuum-continuum delay \(\tau_{cc}\), where used for reference, is evaluated from the `long-range' analytic expression \cite{Dahlstrom2012}.
The comparison with RMT~\cite{RMT} is performed with the same HF model of the atom. The calculation involved a combination of three fields with cos\textsuperscript{2} envelopes: 16-cycle dressing field with central wavelength 800~nm and peak intensity \(10^{11}\)~W/cm\textsuperscript{2}, and the harmonic pair ($\Omega_{13}$ and $\Omega_{15}$) each composed by  240-cycle XUV fields with peak intensity \(10^{10}\)~W/cm\textsuperscript{2}. Both pulses have a duration of approximately 20~fs full width at half maximum (FWHM). The wavefunction was time-propagated for 60~fs in a simulation domain of radius \(r_{\text{max}} \approx 4600~a_0\). 
Fourier filtering is applied to the simulated photoelectron spectra to extract the parameters of Eq.~\eqref{RabbittFormula}.

Angularly resolved interference signal is the most sensitive probe of the relative magnitudes and phases of individual contributing partial waves; it contains enough information for reconstruction of phases of the individual partial waves in two-photon ionization~\cite{He2024}.
\begin{figure}[!htbp]
    \centering
    \includegraphics[width=\columnwidth]{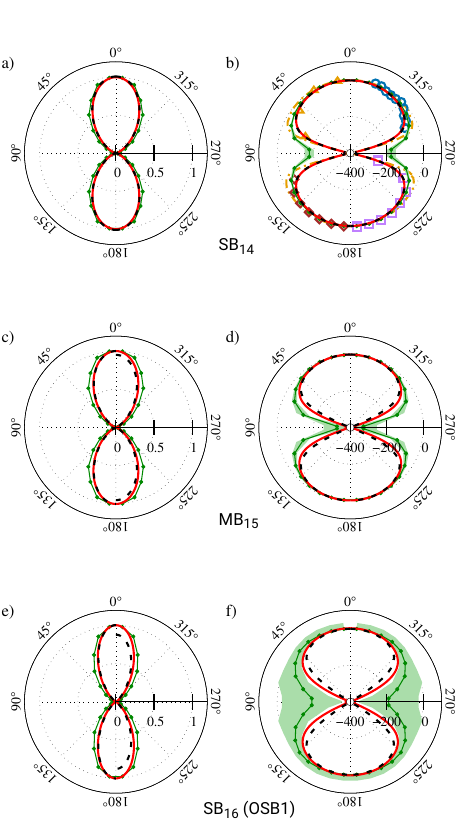}
    \caption{Angularly resolved \Rabbitt{} oscillation amplitude $\mathcal{B}$ (a,c,e) and delay $\tau_R$ (b,d,f) in attoseconds extracted from two-harmonic ($\Omega_{13}$ and $\Omega_{15}$) \Rabbitt{} in argon $3p$ for the channels SB$_{14}$ (a,b), MB$_{15}$ and OSB$_{16}$ as a function of the angle $\vartheta$ between the polarization and photoelectron emission axis. 
    Red solid curves mark the multiphoton results, dashed black represents the partial-wave asymptotic approximation, green with circles is our measurement, green shaded area its experimental uncertainty at 1$\sigma$. The light dash-dot line marks the calculation from~\cite{Jiang2022}. Triangles~\cite{Jiang2022}, empty circles and diamonds~\cite{Cirelli2018} and squares~\cite{RabbittErrorEvaluation} mark earlier measurements at SB\textsubscript{14}. The oscillation amplitudes are normalized to one, except for the black dashed curve in the right half of panels a,c,e, where it is scaled by the same factor as the multiphoton (solid) result. The delays are shifted to zero along the polarization axis for better comparison.}
    \label{fig:ar-angular}
\end{figure}
Figure~\ref{fig:ar-angular} compares the angularly resolved \Rabbitt{} in SB, MB and OSB1 for ioinization from the (complete) \(3p\)-shell. 
The oscillation amplitudes are given by \(\mathcal{B}\propto|T_<^*T_>|\), and the delay by \(\tau_R=\arg(T_<^* T_>)/2\omega\)~\cite{ertel2024a}. 
Our static exchange model is compared with our measurements and also with other results available in the literature \cite{Jiang2022, Cirelli2018, RabbittErrorEvaluation} on the SB\textsubscript{14}. 
It appears that the angularly resolved $\tau_R(\vartheta)$ remains close to zero up to $\vartheta\sim$45$^\circ$ and rapidly decreases to large negative values for angles closer to perpendicular emission.
The distributions of $\tau_R(\vartheta)$ are not exactly the same for the SB, MB and OSB1. The partial waves involved in SBs and MBs do not have the same parities (see Fig.~\ref{IonizationTimeDelayPrinciple}) and cannot reproduce the exact same angular distribution (except if $\tau_R$ is isotropic). The OSB$n$ involves higher angular quantum numbers and concentrates the oscillation along the polarization axis.

\begin{figure*}[!ht]
    \centering
    \includegraphics[width=\textwidth]{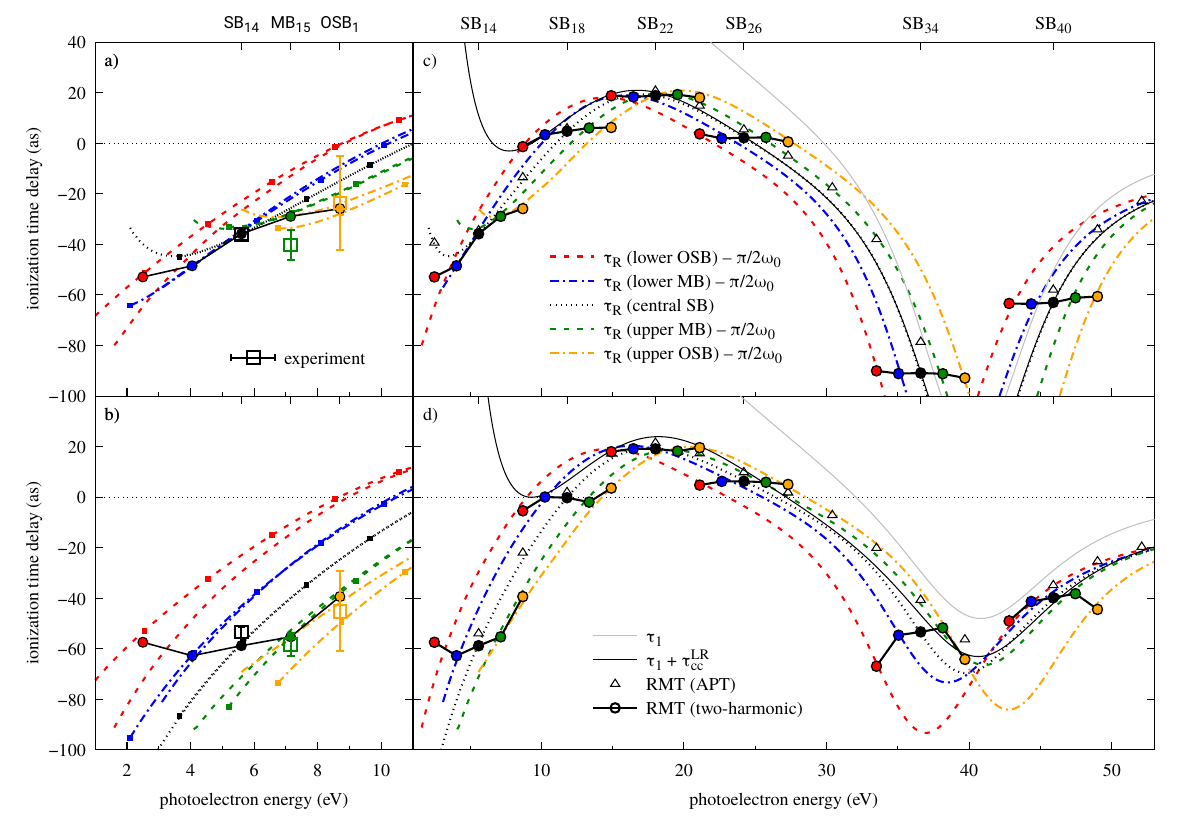}
    \caption{Comparison of the ionization time delay determined from single XUV photon ionisation (solid line, \(\tau_1 + \tau_{cc}^{LR}\)), and the predicted oscillation phase along the polarisation axis for the SB, MB and OSB (broken lines, \(\tau_R\)) at a given final photoelectron energy, all from time-independent lowest order perturbation theory. Connected circles (labeled `RMT (two-harmonic)`) represent delays extracted from six separate non-perturbative time-dependent simulations of `two-harmonic \Rabbitt{}', each giving rise to a specific sideband (labelled atop) and to the surrounding bands. The two methods are in very close agreement. Empty triangles show a time-dependent simulation with a full comb attosecond pulse train.  Emission (a,c) along polarization axis and (b,d) angularly-integrated. Panels (a) and (b) show a subsets of (c) and (d) with partial-wave asymptotic results included as lines with squares and compared to experiment. The experimental points were uniformly shifted to match the calculation for oriented emission at SB\textsubscript{14}.}
    \label{SB_Oscillations}
\end{figure*}

Figures~\ref{fig:ar-angular}(a,c,e) show that the angularly resolved oscillation amplitude $\mathcal{B}$ is almost the same for SB and MB despite a different arrangement of dressing photons (see Fig.~\ref{RabbittPrinciple}(a,b)) and gets narrower with the number of dressing photons involved, such as for OSB1. 
This can be  understood using the asymptotic approximation, even when the angular momentum dependence of the free-free transition coefficient is neglected. As there are no ion-ion transitions in the static exchange model of argon, the amplitude of a higher-order \Rabbitt{} ionization for oriented emission can be written as
\begin{equation}
    T_\textrm{fi}^{(n + 1)}(\bm{k}) \sim (\hat{\bm{\varepsilon}}_{\text{IR}} \cdot \hat{\bm{k}})^n A_{\kappa_{n - 1} k}^{(1)} \dots A_{\kappa \kappa_1}^{(1)} \dd_\textrm{fi}^{(1)}(\hat{\bm{k}}\kappa) \,.
    \label{eq:Tfin-universal}
\end{equation}
This high-energy limit suggests that the \Rabbitt{} signal
\begin{equation}
    \mathcal{B}(\bm{k}) \sim |T_\textrm{fi}^{(p)*}(\bm{k}) T_\textrm{fi}^{(q)}(\bm{k})| \,,
\end{equation}
where $p$ and $q$ represent the number of involved photon in each path,
acquires an additional emission-angle-dependent factor \(\cos^2\vartheta = (\hat{\bm{\varepsilon}}_{\text{IR}} \cdot \hat{\bm{k}})^2\) for each absorbed \(\omega\)-photon. This factor is responsible for making the distribution more peaked along the polarization axis. One can ascribe this cosine factor to a simple geometrical effect of projection of the electric strength vector along the photoelectron emission direction. In other words, the leaving photoelectron is, asymptotically, affected only by a parallel component of the dressing field.

As established by Bharti \textit{et al.}~\cite{Bharti2021}, if the approximation given by Eq.~\eqref{eq:Tfin-universal} was accurate, the coefficients \(A_{kk'}^{(1)}\) in each of the two \Rabbitt{} ionization pathways would combine to the universal continuum-continuum delays in such a way that the measured \Rabbitt{} delay would be the same when extracted from any of the bands or sidebands arising from the same pair of harmonics. The only difference predicted by the standard asymptotic theory would be an overall phase factor \( i^{-p + q}\) arising from unbalanced number of \(i\) factors in the products of the \(A_{kk'}^{(1)}\) coefficients present in each pathway, see Eq.~\eqref{eq:Akk}. This factor is responsible for the \(\pi\)-shift of the other bands with respect to the central sideband. However, both the theory and the experiment confirm that distinct bands exhibit slightly different angular distribution of the delays, particularly at low energies. This points to insufficiency of the standard asymptotic approximation, Eq.~\eqref{eq:Tfin-universal}, and to a need for the more complete, partial-wave-sensitive theory.

Figure~\ref{SB_Oscillations}  presents simulated time delays along the polarization axis (a and c) and integrated angularly (b and d) on the low photon energy range of interest (a,b) and  over a wider energy range (c,d).
The angularly integrated case corresponds to the widest available \Rabbitt{} measurements performed with a simple electron spectrometer such as magnetic bottle.
The delays obtained from the emission-integrated ionization signal is described as,
\begin{align}
    I(\tau) &\sim \int T_\text{fi}^{(p)*}(\bm{k}) T_\text{fi}^{(q)}(\bm{k}) d^2\hat{\bm{k}} \nonumber \\
    &\sim \int \mathcal{B}(\bm{k}) e^{2i\omega\tau_R(\bm{k})} d^2\hat{\bm{k}}\,.
\end{align}

Along the polarization axis, Fig.~\ref{SB_Oscillations}(a), the predicted $\tau_R$ is almost the same for the SB, MB and OSB specially above 10~eV. In practice, it corresponds to a $\pi$-shift in Eq.~\eqref{RabbittFormula} as predicted under the soft photon approximation (see Supplementary Material~\cite{supp}). The very good agreement between predicted delays extracted from different orders of \Rabbitt{} is caused by the nearly identical \(A_{\kappa \lambda kl}\) factors pertaining to different partial waves at energies greater than 10~eV, together with absence of nodes in relevant spherical harmonics for the axial emission. Consequently, the multiphoton amplitude can be factorized and the \Rabbitt{} delay is then well-separable into the asymptotic components according to Eq.~\eqref{eq:Tfin-universal}.

However, the non-trivial angular dependence of $\mathcal{B}(\vartheta)$ and $\tau_R(\vartheta)$, which is also different for the individual higher-order \Rabbitt{} pathways, is responsible for slight differences in the integrated time delay extracted from the individual bands, Fig.~\ref{SB_Oscillations}(b). Such differences fall in the order of tens of attoseconds for argon in the range under consideration.

In general, the delays extracted from the mainbands are very similar to those obtained from the central sidebands even in the emission-integrated case, but this is not always the case with delays obtained from the outer sideband. Such behaviour is easy to explain within the standard asymptotic theory and it is the consequence of the angular integrals appearing in construction of higher orders. The phase \(\phi_{2\omega} = 2\omega\tau_R\) of the emission-integrated \Rabbitt{} signal in the central sideband is
\begin{align}
    \phi_{2\omega}^\text{SB} &\approx 
    \arg \sum_{l'm'lm}
    \langle l'm'|(\hat{\bm{\varepsilon}}_\text{IR} \cdot \hat{\bm{r}})^2 |lm\rangle
    d_{\text{f}l'm',\text{i}}^{(1)*}(\kappa_+) d_{\text{f}lm,\text{i}}^{(1)}(\kappa_-) \nonumber \\
    &+ \arg A_{\kappa_+ k}^{(1)*} A_{\kappa_- k}^{(1)} \,,
    \label{eq:phi14}
\end{align}
while for the upper mainband (MB$_>$) one gets
\begin{align}
    \phi_{2\omega}^{\text{MB}_>} &\approx 
    \arg \sum_{l'm'lm}
    \langle l'm'|(\hat{\bm{\varepsilon}}_\text{IR} \cdot \hat{\bm{r}})^2 |lm\rangle
    d_{\text{f}l'm',\text{i}}^{(1)*}(\kappa_+) d_{\text{f}lm,\text{i}}^{(1)}(\kappa_-) \nonumber \\
    &+ \arg A_{\kappa_+ k}^{(1)*} A_{k \kappa_-}^{(1)*} \,,
    \label{eq:phi15}
\end{align}
which differs only in the continuum-continuum factors and these are identical, up to a sign factor, due to complex-conjugation properties of Eq.~\eqref{eq:Akk}. In contrast, the phase of emission-integrated OSB1 results in
\begin{align}
    \phi_{2\omega}^\text{OSB1} &\approx 
    \arg \sum_{l'm'lm}
    \langle l'm'|(\hat{\bm{\varepsilon}}_\text{IR} \cdot \hat{\bm{r}})^4 |lm\rangle
    d_{\text{f}l'm',\text{i}}^{(1)*}(\kappa_+) d_{\text{f}lm,\text{i}}^{(1)}(\kappa_-) \nonumber \\
    &+ \arg A_{\kappa_+ \kappa}^{(1)*} A_{\kappa \kappa_-}^{(1)*} A_{\kappa_- k}^{(1)*} A_{\kappa_- k}^{(1)} \,,
    \label{eq:phi16}
\end{align}
which, at the level of Eq.~\eqref{eq:Akk}, yields the same continuum-continuum phase (second term of Eq.~\eqref{eq:phi16}), again, because the last two \(A\)-factors cancel in phase, but the one-photon part (first term of Eq.~\eqref{eq:phi16}) is different in Eqs.~\eqref{eq:phi14} or~\eqref{eq:phi15}. In energy regions dominated by a single partial wave (\(d\)-wave in the present case around SB\textsubscript{22}) the partial wave mixing by the angular integral is unimportant and the delays obtained from SB (Eq.~\eqref{eq:phi15}) and OSB1 (Eq.~\eqref{eq:phi16}) are in agreement.

Only for the case of the lowest sideband SB$_{14}$ and for both the axial case and the angle integrated case the neighboring inner and outer sidebands do not reproduce the delay given by the central sideband. This is caused by the strong energy dependence of the free-free transitions close to the threshold, while our asymptotic theory reproduces this effect accurately. The excellent agreement with our experiment confirms this effect.

\section{Practical implementation}
\label{sect:practical}

In practice, reconstruction of higher-order ionization amplitudes (at the leading order of perturbation) follows the scheme indicated in Fig.~\ref{IonizationTimeDelayPrinciple}(c) and Eqs.~\eqref{eq:T2}--\eqref{eq:T2ion}. Specifically, in absence of permanent dipole moments and of near-resonant dipole transitions within the residual ion the application of the partial-wave asymptotic approach reduces to the following steps: 

\begin{itemize}
\item[1.] The partial-wave-resolved one-photon XUV ionization amplitudes \(\tilde{T}_{\pm,\text{fi},lm}^{(1)} := 2\pi i d_{\text{f}lm,\text{i}}^{(1)}(\kappa_\pm)\) defined in Eq.~\eqref{eq:dfi} are calculated for intermediate photoelectron momenta \(\kappa_\pm\).
\item[2.] All partial-wave amplitudes are multiplied by the angular coupling coefficient and by the continuum integral coefficient describing the free-free transition,
\begin{equation}
    \tilde{T}_{\pm,\text{fi},lm}^{(n+1)} =
    -\sum_{\lambda\mu} \langle lm | \hat{\bm{\varepsilon}}_{\text{IR}}\cdot \hat{\bm{r}} |
    \lambda\mu \rangle
    A_{\kappa_\pm\lambda kl}^{(1)}
    \tilde{T}_{\pm,\text{fi},\lambda\mu}^{(n)} \,,
    \label{eq:Tn+1}
\end{equation}
to produce amplitudes for the higher order at the photoelectron momentum \(k = (\kappa_\pm^2 \pm 2\omega)^{1/2}\); cf.\ Eq.~(C9) of~\cite{Benda2022}. For a given polarization \(\hat{\bm{\varepsilon}}_\text{IR}\) the angular integral in Eq.~\eqref{eq:Tn+1} can be expressed using Clebsch-Gordan coefficients. The radial integral \(A_{\kappa_\pm \lambda kl}^{(1)}\) is plotted in Fig.~\ref{fig:akkl} and its precomputed values for \(l \le 10\) and \(k^2/2 \le 150\)~eV are available in the Supplementary Material~\cite{supp}.
If needed, this step is repeated for each additional absorbed IR photon with a correspondingly increasing (or decreasing) final photoelectron momentum.
\item[3.] The final multiphoton partial-wave amplitudes are multiplied by the combinatoric factor,
\begin{equation}
    T_{\pm,\text{fi},lm}^{(n)} = \frac{1}{(n - 1)!} \tilde{T}_{\pm,\text{fi},lm}^{(n)} \,,
\end{equation}
and used to evaluate the observables of interest, for instance the higher-order emission-integrated \Rabbitt{} delay
\begin{equation}
    \tau_R = \frac{1}{2\omega} \arg \sum_{lm} T_{+,\text{fi},lm}^{(p)*} T_{-,\text{fi},lm}^{(q)} \,.
\end{equation}
\end{itemize}
If the residual ion has a permanent dipole or supports dipole transitions, the second step needs to be supplemented with analogical contribution from the ion-ion terms, Eqs.~\eqref{eq:T2ion} and~\eqref{eq:Akk0}.

\section{conclusion}
\label{sect:conclusion}

In this work, we provide a numerically cheap and accurate solution to take into account the dressing photons in a \Rabbitt{} experiment. It builds on the well-accepted asymptotic approximation theory, but represents the interaction of the photoelectron with IR field separately for each partial wave. This allows us to take into account the different ionization phase acquired by distinct partial waves. Unequal phases from the IR absorption imprinted to different partial waves then affect reconstruction of the angularly resolved \Rabbitt{} oscillation, where multiple partial waves interfere, as well as orientation-unresolved measurements around Cooper minima and similar interference structures. The method can be applied to the emerging  \Rabbitt{} variants where the dressing field corresponds to the energy difference between the harmonics \cite{laurent2012PRL, Loriot2020}, the third \cite{Nature_Rabbitt3w} or the fourth \cite{Bharti2023}.

We implemented a  two-harmonic  \Rabbitt{} experimental setup to isolate the interference pathways and observe the higher orders free of overlap between the contributions. This allows the disentaglement of various ionization pathways as well as investigation of individual multi-photon interferences in higher-order \Rabbitt{} processes. The method is particularly promising for applications involving high density of electronic states such as in molecules. We have shown that even in the emission-integrated case the delays obtained from MBs are very accurately in phase opposition to the central SB, in agreement with the `rule of thumb'~\cite{Bertolino2021}. The OSBs are similar in this regard, but they are affected slightly differently by the integration over emission directions if multiple partial waves contribute comparably, which may affect the extracted delays close to Cooper minima and similar interference structures.

We have shown that the new theoretical method gives valid results in argon even at very low energies, suggesting that it is particularly the effect of the angular momentum barrier that is the most lacking in the traditional asymptotic formalism of laser-assisted ionization. We also analyzed the higher-order \Rabbitt{} schemes, where a similar partial-wave asymptotic factorization is performed for the higher-order photoionization amplitudes. This also gives, asymptotically, results that match the full leading-order perturbation theory amplitudes, but at low energies the accuracy of the approximate treatment is reduced. We demonstrated that the partial-wave asymptotic approach provides similar results compared to time-dependent simulations and the time-independent leading-order perturbation method.

Finally, the method is compatible with any computational approach that provides one-photon ionization amplitudes in the partial-wave decomposition. The results are very accurate for settings where electron correlation in the continuum is negligible. When it is not, the photoelectron in the vicinity of the molecule exchanges energy with multiple ionic states and does not have a well-defined energy. Consequently, approximating the short-range part of the photoelectron wavefunction with a pure Coulomb wave in Eq.~\eqref{eq:Tlli} is no longer justified. Such a situation may arise for example in case of the strongly dipole-coupled B~\(^2\Sigma_u^+\) and C~\(^2\Sigma_g^+\) states of CO$_{2}{}^{+}$~\cite{Benda2022} and in the core-excited shape resonance in C \(^2\Sigma_g^+\) state which couples to hundreds of ionic states~\cite{harvey2014,masin2018a}. This seeming deficiency can be actually employed as a tool to identify such correlated systems. Whenever the predictions of the partial-wave asymptotic approach diverge from experiment or full calculation at low energies, it is a signature of strongly coupled channels.



\begin{acknowledgments}
This work has been supported by the Charles University Research Centre program No.\ UNCE/24/SCI/016. Computational resources were provided by the e-INFRA CZ project (ID:90254), supported by the Ministry of Education, Youth and Sports of the Czech Republic. Zdeněk Mašín and Jakub Benda acknowledge the support of the Czech Science Foundation (Grant No. 20-15548Y).
We acknowledge support from the CNRS, ANR-21-CE30-0052 ‘FAUST’, the F\'ed\'eration de recherche Andr\'e Marie Amp\` ere and the European COST Action AttoChem (CA18222). We acknowledge Evan Langloÿs for experimental support.
\end{acknowledgments}

\appendix

\bibliography{Main.bib}
\end{document}